\documentstyle[prb,aps]{revtex}
\begin{document} 
%\draft
\title{Vibrations of a chain of Xe atoms in a groove of carbon nanotube
bundle}

\author{Marko T. Cvita\v{s}\thanks{E-mail: m.t.cvitas@durham.ac.uk}}
\address{Department of Chemistry, University of Durham, South Road, Durham
DH1 3LE, UK}
\author{Antonio \v{S}iber\thanks{E-mail: asiber@ifs.hr}}
\address{Institute of Physics, P.O. Box 304, 10001
Zagreb, Croatia
}
\maketitle

\begin{abstract}
We present a lattice dynamics study of the vibrations of a linear
chain of Xe adsorbates in groove positions of a bundle of carbon
nanotubes. The characteristic phonon frequencies are
calculated and the adsorbate polarization vectors discussed.
Comparison of the present results with the ones previously published
shows that the adsorbate vibrations cannot
be treated as completely decoupled from the vibrations of carbon
nanotubes and that a significant hybridization between the adsorbate and
the tube modes occurs for phonons of large wavelengths.
\\
\begin{center}
{\bf Published in Phys. Rev. B 67, 193401 (2003)}
\end{center}
\end{abstract}
%%%!!!!! CHECK PACS !!!!

Adsorption of gases in carbon nanotube materials has recently attracted 
considerable
attention. \cite{MWCPRL,uptake,Calbi1,Colpriv,sibphon,sibgroove}
Studies of this phenomenon are in part
motivated by the suggested application of carbon nanotubes for an
efficient gas storage. \cite{Dillon} Theoretical studies of adsorption
in carbon nanotubes have
predicted unique, one-dimensional (1D) phases of the adsorbed
gases. \cite{MWCPRL,Calbi1,sibphon} The dimensionality of the adsorbate
structures is reduced due to a specific geometry of an isolated carbon
nanotube and bundles (or ropes) of carbon nanotubes. This opens a
possibility to use nanotubes as a template for growth of adsorbed
nanostructures such as atomic and molecular "wires". \cite{GaoR}

The nanotube materials commonly studied and produced do not contain
isolated carbon nanotubes, but the so-called bundles of nanotubes,
containing $\sim$ 100 nanotubes organized in a triangular
lattice.\cite{Ayan} The
bundle surface can support 1D phases of matter. \cite{Calbi1} The
potential for adsorption of gases on the bundle surface is
stronger in the region between the two tubes (groove) than in other
positions. \cite{sibgroove} 
Thus, the atoms bound in these positions should desorb from the surface 
at higher temperatures than those adsorbed at other places, which 
clearly suggests that the 1D phases of matter adsorbed in the 
groove positions can be prepared by carefully programmed thermal
desorption.

The detection of such phases by specific heat measurements has been
suggested in Ref. \onlinecite{sibphon} and their signature
in the low-temperature specific heat has been theoretically predicted. 
The approach of Ref. \onlinecite{sibphon} was based on two simplifying 
assumptions: {\em (i)} the discreteness of the carbon nanotube was
neglected and {\em (ii)} the nanotubes were treated as completely
rigid, providing a static external potential for the adsorbed atoms. The
adsorbate vibrational modes obtained in that study were 
completely localized on the adsorbate atoms due to the neglect of coupling 
between the adsorbate and substrate vibrations. As the
low-temperature thermodynamics of adsorbates crucially depends on the
details of the excitation spectrum, more realistic model of
adsorbate vibrations is needed in order to predict and interpret correctly 
measurements of thermodynamical quantities such as specific heat.

In this article, we present calculations of the phonon
dispersions and phonon eigenvectors of a 1D phase of Xe atoms adsorbed in
a groove of carbon nanotube bundle. 
Unlike in the previous studies focused on adsorption
of gases\cite{MWCPRL,uptake,Calbi1,Colpriv,sibphon,sibgroove}, we
do not treat the nanotubes as rigid and smooth objects which enables us 
to examine the effects of hybridization of adsorbate and substrate
vibrations and to calculate the phonons of the realistically modeled
system.

Due to a large radius of Xe atoms the occupation of
interstitial channel positions is not possible. For nanotubes with closed
ends, the adsorption in the tubes is also not possible. Thus, for low
coverage and
pressure, only the groove positions are expected to be occupied.
The system we examine is represented in Fig. \ref{fig:fig1}. It consists
of two infinitely long (17,0) single wall carbon nanotubes with a Xe chain
in between them. The two tubes are mirror images of one another and
their axes are separated by $2R + 3.2$\AA, where $R$ is the tube radius
[$R=6.6$ \AA \hspace{0.8mm} for (17,0) tube; note that the diameter of
(17,0) nanotube is close to mean diameter of carbon nanotubes in the
samples studied experimentally\cite{Rols1}].
For this particular, highly symmetric system, Xe phase commensurate 
with the tubes is expected to form in a range of temperatures, similarly
as for Xe monolayers on graphite surface. \cite{Bracco1} The
commensurability of this phase is the main reason for its choice in the
present study, since the periodicity of its structure simplifies the
calculations. Such a phase
is unlikely to exist
in a realistic situation, where a range of finite axial and angular
offsets of the two neighboring tubes are to be
expected. Nevertheless, the coupling between the adsorbate and substrate
vibrations persists even in an
incommensurate adsorbate phase\cite{Black}, which may be expected in the
case of a poor alignment of the adjacent two tubes. This
coupling can be easily studied for a commensurate phase represented
in Fig. \ref{fig:fig1} where a lattice
dynamics approach\cite{deWette} is
possible. The general features of the coupling between the adsorbate and
substrate vibrations are expected to be similar in incommensurate and
commensurate phases, based on previous
experimental and theoretical evidence for such phases of adsorbate
monolayers on crystal
surfaces. \cite{Black,APGrah,SibXePR}. The fact that we consider
only two tubes defining a groove region is not of importance, since the
interaction of Xe atoms with more distant tubes in the bundle
is very small due to large Xe-C distances for C atoms forming these
nanotubes.

The nearest neighbor Xe-Xe distance in the phase depicted in
Fig. \ref{fig:fig1} is $3a=4.26$ \AA, where $a$ is the nearest neighbor
C-C spacing. The position of a single Xe adsorbate
was found by the potential energy minimization procedure, using 
Xe-C binary potential suggested in Ref. \onlinecite{uptake}, also used
in Ref. \onlinecite{sibphon}. As a result of this procedure, we 
obtained that the height of the Xe chain above the plane 
passing through the tube axes is 6.3 \AA. The structure we study can be
obtained by one-dimensional discrete translations of a unit cell
containing 136 C and one Xe atom. The vibrations of isolated carbon
nanotube were modeled using a set of force constants for
intraplanar interactions in graphite
developed in Ref. \onlinecite{Smith}. As noted by Lu\cite{Lu}, the
intertube interactions must be given a special treatment, and for these
interactions we have adopted the approach of Ref. \onlinecite{Lu}, i.e. we
introduced an effective
binary C-C potential of Lennard-Jones (LJ) form with the parameters
$\epsilon_{C-C}=12$ meV and $\sigma_{C-C}=3.4$ \AA. Xe-Xe interactions
were modeled by LJ pair potential with parameters
$\epsilon_{Xe-Xe}=19.07$ meV and $\sigma_{Xe-Xe}=4.1$ \AA. \cite{uptake}
More refined Xe-Xe potentials exist\cite{HFDB2}, 
but we choose the one employed in Ref. \onlinecite{sibphon} in order to
examine the effects of approximations introduced in that work.
The results of the calculation with more
precise Xe-Xe potential will be discussed later.

The calculated phonon dispersions, $\omega(q)$, where $\omega$ and $q$ are
the phonon frequency and wave vector, and the corresponding density of
states, $\rho(\omega)$, are presented in Fig. \ref{fig:fig2}. The
density of states is calculated as
\begin{equation}
\rho(\omega) = \sum_{i} \sum_{q} \delta[\omega - \omega_i (q)],
\label{eq:dens1}
\end{equation}
where the index $i$ labels the different phonon branches.
In plotting this quantity, we have represented the $\delta$-functions in
Eq. (\ref{eq:dens1}) with normalized Gaussian functions with width
parameters of 0.3 meV [panel (b)] and 0.15 meV [panel (d)]. The
high energy portion of the phonon
spectrum (panels (a) and (b) of Fig. \ref{fig:fig2}) correlates quite well
with the spectrum of isolated nanotube studied in
Refs. \onlinecite{tube1,tube2,tube3}. At
frequencies between 0.8 and 6.0 meV, the adsorbate vibrations dominate the
phonon density of states and the peak in the density of states (denoted by
arrow in panel (b) of Fig. \ref{fig:fig2}), not originating from the tube
vibrations is clearly seen. The low frequency region of the dispersion
curves and the density of states is shown in panels (c) and (d) of
Fig. \ref{fig:fig2}. The three adsorbate induced modes denoted by $L,T1$
and $T2$ are clearly visible. To examine the polarization of these modes, 
we plot in Fig. \ref{fig:fig3} the quantity defined as
\begin{equation}
\rho_{\nu}^{Xe} (\omega, q) = \sum_{i} |e_{i,\nu}^{Xe}(q)|^2 \delta
[\omega - \omega_i (q)], \, \nu=x,y,z,
\label{eq:dens2}
\end{equation}
where $e_{i,\nu}^{Xe}(q)$ is the $\nu$-th Cartesian component of the
polarization vector of the $i$-th phonon branch at the
position of a Xe atom. The quantity in Eq. (\ref{eq:dens2}) is
directionally
and wave vector resolved density of Xe vibrations.

One can observe in Fig. \ref{fig:fig3} that the mode
denoted by $L$ corresponds to vibrations of the Xe chain polarized along
$x$ direction i.e. along the chain direction (longitudinal mode), while
the modes denoted by $T1$ and $T2$ correspond to vibrations of Xe atoms in
$y$ and $z$ directions, respectively (transverse modes). The dashed lines
in panel (c) of Fig. \ref{fig:fig2} represent the dispersions of L,T1 and
T2 modes obtained in Ref. \onlinecite{sibphon}. Note that
in contrast to Ref. \onlinecite{sibphon} the $L$-mode acquires a gap
($\sim$ 0.8 meV) at the
center of the quasi one dimensional Brillouin zone which is a
consequence of the commensurability of
the Xe chain with the substrate. A similar effect has been observed for
commensurate Xe
overlayers on crystal surfaces.\cite{SibXePR,XePt,ToennVolm}
While the
$L$ mode does not hybridize with the tube modes, the two transverse
adsorbate modes are influenced by the elasticity of the tube
substrate which is clearly seen for wave vectors smaller than about 0.1
1/\AA. In this long-wavelength region, the adsorbate atoms vibrate with
several frequencies determined by the details of the nanotube phonon
spectrum. The vibrations of Xe atoms in this region cannot be
treated as vibrations in which {\em only} Xe atoms participate, since the
atoms of the two substrate tubes are also involved in these vibrations. As
a result, the localization of these modes on Xe atom is
smaller from the corresponding value at the end of the Brillouin zone
($q=0.737$ 1/\AA), where the $L,T1,$ and $T2$ modes represent 100 \% pure
Xe vibrations in which the substrate atoms do not participate. The
adsorbate-substrate mode mixing appears only in the long wavelength region
due to large "stiffness" of the nanotube phonons when compared to
adsorbate modes. Thus, the modes cross only for small q values.

Examination
of the polarization vectors of all atoms in the unit cell reveals that the
motion of carbon atoms in the region of hybridization is highly
complex. The displacements of C atoms from their equilibrium positions
have in general all three Cartesian components. Transforming these
components into cylindrical coordinate
system does not lead to much more insight since the modes are again highly
complex and have radial, tangential and $z$-components which are all
different from zero. However, it can be said that the radial components of
the polarizations vectors of C atoms slightly dominate the motion pattern.

We find that the frequencies of
the two transverse phonons agree reasonably well with the results
published in Ref. \onlinecite{sibphon}. The
differences can be attributed to the facts that in
Ref. \onlinecite{sibphon} the two tubes were treated as smooth
and that the tubes considered in Ref. \onlinecite{sibphon} were of
somewhat larger radius (6.9 \AA) than the ones studied in this work
(6.6 \AA). As in Ref. \onlinecite{sibphon}, we find that the
adsorbate transverse vibrations polarized in the $z$-direction exhibit
lower frequencies than those polarized in the $y$-direction. This is a
consequence of a specific shape of the potential experienced by Xe
adsorbate in the groove (Fig. 2 of Ref. \onlinecite{sibphon}).

The maximum frequency of the $L$-mode obtained
in this study is almost twice larger than the one obtained in
Ref. \onlinecite{sibphon}. The reason for this is that the
nearest neighbor Xe-Xe radial force constant is significantly larger in
the commensurate phase than in the incommensurate phase studied in
Ref. \onlinecite{sibphon}. This is due to smaller Xe-Xe distances dictated
by the corrugation of the underlying substrate (bundle of tubes). 

Use of more realistic Xe-Xe potentials\cite{HFDB2} results in lower
frequency of the $L$-mode at the
Brillouin zone boundary. Using the HFDB2 Xe-Xe potential
suggested in Ref. \onlinecite{HFDB2}, we obtain the $L$-mode
frequency at the zone edge of 4.62 meV, while the $L$-mode zone center gap
does not change (0.8 meV). The two
transverse modes retain practically the same dispersions as those depicted
in Fig. \ref{fig:fig2}. This is not surprising, since the transverse mode
frequency is determined mainly by Xe-C interactions. Renormalization of
the Xe-Xe interactions by the substrate\cite{kostov1,Bruchbook} results in
a modified effective Xe-Xe potential. In the study of these effects
reported in Ref. \onlinecite{kostov1}, it was found that the interaction
between the two Xe atoms adsorbed in the groove positions is weaker than
in free space and that it can be approximately represented
by a scaling of the energy parameter of the LJ Xe-Xe
potential.\cite{kostov1} Applying an analogous scaling to the HFDB2
potential, we obtain further reduction of the maximum $L$-mode frequency,
which now amounts to 4.05 meV (at the zone edge).

It is of interest to compare the present results with the theoretical and
experimental results for vibrations of the commensurate ($\sqrt 3 \times
\sqrt 3)$R30$^o$ phase\cite{Bracco1} of monolayer of Xe atoms on
graphite. The arithmetic mean of the zone edge
frequencies of the two transverse modes in the present system is quite
close to the measured frequency of
the transverse mode (with polarization perpendicular to the surface) of
a Xe monolayer on graphite\cite{ToennVolm} ($\approx$ 3
meV). Molecular dynamics calculations for commensurate monolayer
phase of Xe on graphite\cite{Marchese} yielded $L$-mode dispersion
with maximum frequencies of 5.6 meV (M point in the surface Brillouin
zone) and 4.85 meV (K point in the surface Brillouin zone). These values
are comparable to the maximum frequency of the $L$ mode we obtain in
this work. The zone center gap of $L$-mode was found to be
$\sim 0.9$ meV, also very close to the value we obtained for the Xe chain
phase (0.8 meV). Very similar results have been obtained in the lattice
dynamics study of a Xe monolayer on graphite in
Ref. \onlinecite{Rouf}. Some
points in connection with this comparison should be emphasized,
however. Whereas the two transverse modes are
practically dispersionless in the chain phase we study, the in-plane
transverse vibration of Xe monolayer on graphite exhibits a large
dispersion. \cite{Marchese,Rouf} This is a direct consequence of quite
different geometries of the substrate material in the case of adsorption 
on a planar graphite and on the bundle of carbon nanotubes. Additionally,
although the characteristic phonon frequencies are similar for
Xe/graphite and Xe/groove systems, one has to remember that the
dimensionality of the phonon phase space is different in the two
systems. Wave vector $q$ is a one dimensional quantity in the present
case, whereas in the case of Xe/graphite, it is a two dimensional
quantity. This has
direct consequences on the low-temperature thermodynamics of adsorbed
phases. \cite{sibgroove}

In summary, we have investigated the vibrations of a commensurate phase of
Xe atoms adsorbed in the groove positions of a carbon nanotube
bundle. Model we used is significantly improved over previous models
which have treated the nanotube substrate as rigid and
smooth.\cite{MWCPRL,uptake,Calbi1,sibphon,sibgroove} This
enabled us to study the details of the adsorbate phonon spectrum and we
have shown that a "mixing" between the transverse adsorbate
(Xe) and substrate (nanotube bundle) phonon modes can be expected for
large phonon wavelengths. Such mixing does not appear for the adsorbate
mode with longitudinal polarization. Since the low-temperature thermal
properties are dominated by excitations of low-frequency modes (which is
precisely the region where the hybridization occurs), it can
be expected that the signature of the adsorbate phase in the overall
low-temperature thermal properties of the sample will be more complicated
than previously predicted.\cite{Colpriv,sibphon,sibgroove}

\begin{figure}
\caption{
Two (17,0) carbon nanotubes with a commensurate Xe chain in between    
them. This distance ratios in this figure correspond to the ones used
in the calculations. Our choice of the coordinate system is also denoted.
}
\label{fig:fig1}
\end{figure}

\begin{figure}
\caption{
(a) Phonon modes of a system displayed in
Fig. \ref{fig:fig1}. (b) Phonon density of states. (c) Low
frequency portion of the data presented in panel a). Dashed lines
represent L,T1, and T2 modes obtained in
Ref. 5. 
(d) Low frequency portion of the data presented in panel b). The density
of states from panel b) is multiplied by a factor of 3 to obtain the data
in panel d).
}
\label{fig:fig2}
\end{figure}

\begin{figure}
\caption{
Contour-like plot of $x-$ (top panel), $y-$ (middle panel), and $z-$
(bottom panel) components of the directionally and wave vector resolved
phonon density of states at Xe atom. The density of dots in these
plots corresponds to $|e_{i,\nu}^{Xe}(q)|^2$, where $\nu=x$, $\nu=y$, and
$\nu=z$ for the top, middle and bottom panel, respectively. The highest
density of dots corresponds to $|e_{i,\nu}^{Xe}(q)|^2=1$.
}
\label{fig:fig3}
\end{figure}

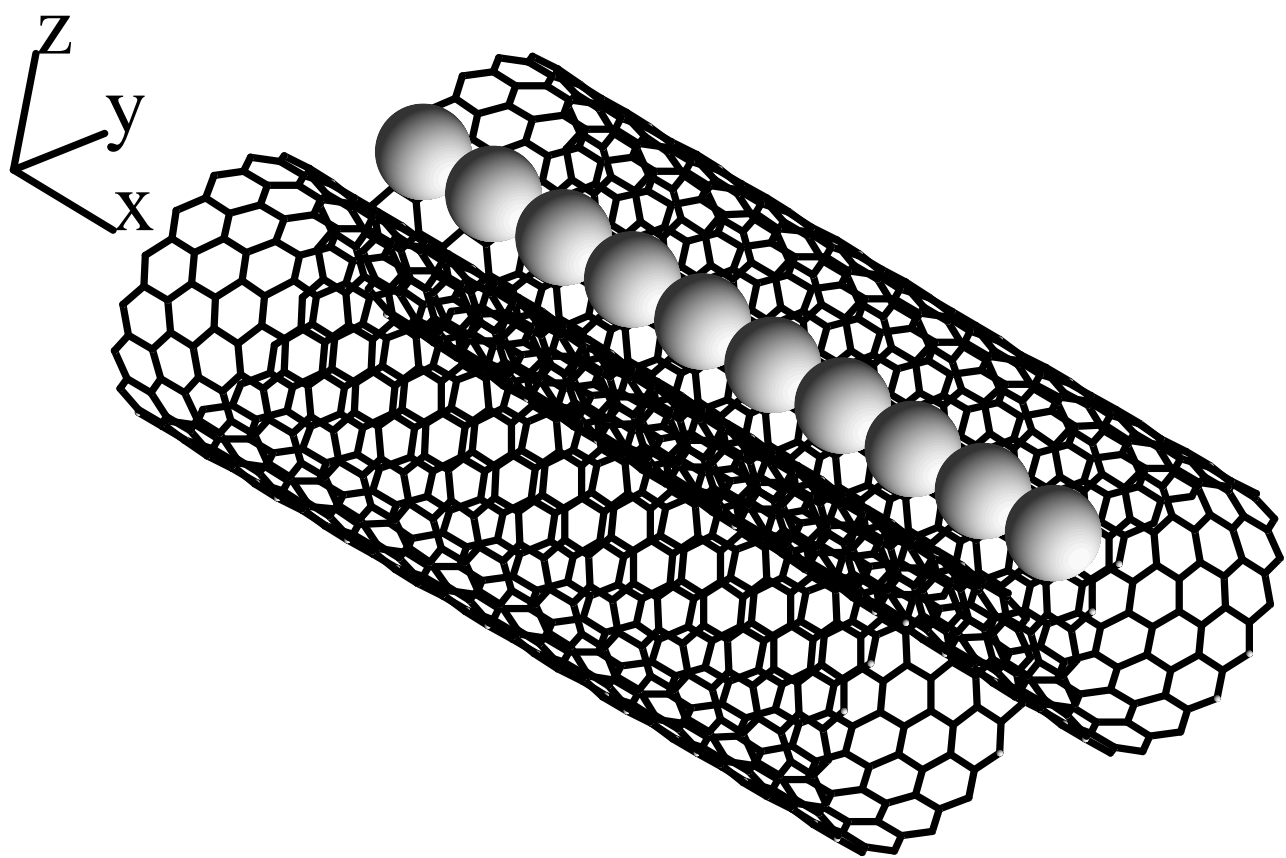
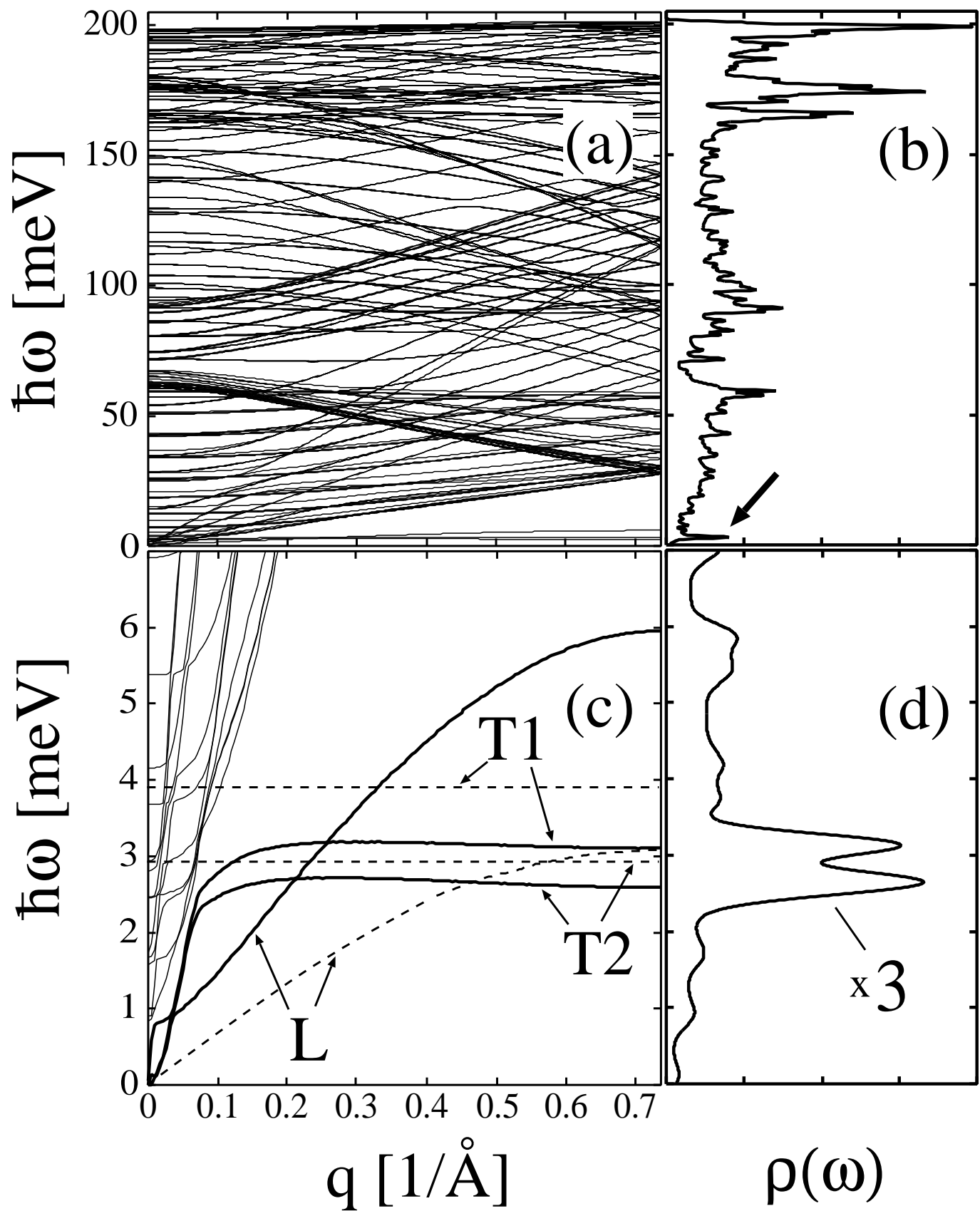
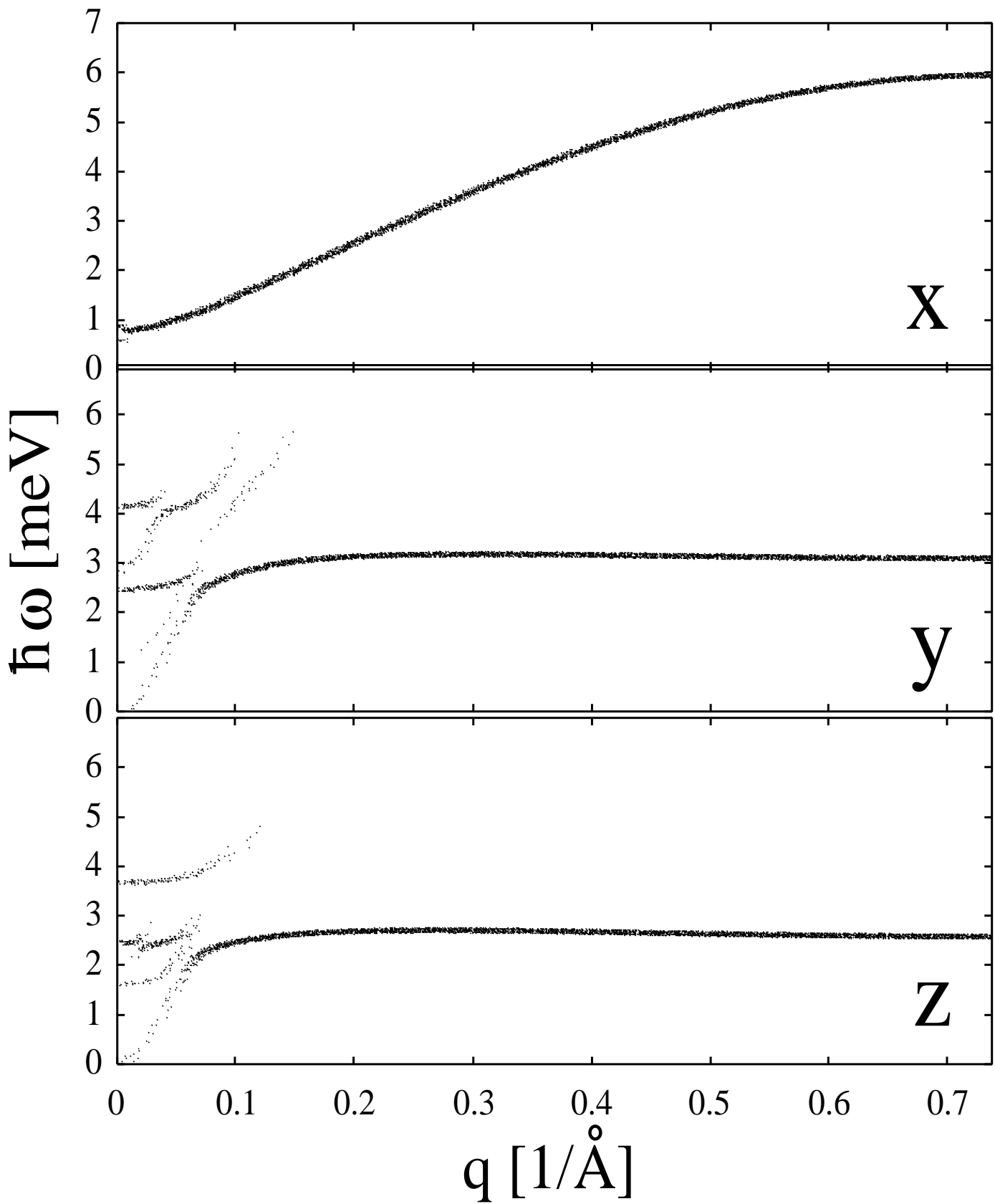
\end{document}